\documentclass[journal=jacsat,manuscript=article]{achemso}

\usepackage[version=3]{mhchem}
\usepackage{bm}

\author{Xinqiang Ding}
\author{Bin Zhang}
\email{binz@mit.edu}
\affiliation
{Department of Chemistry, Massachusetts Institute of Technology, Cambridge, Massachusetts 02139, United States}
\email{binz@mit.edu}

\title{Contrastive Learning of Coarse-Grained Force Fields} 


\usepackage[dvipsnames]{xcolor}
\definecolor{myblue}{RGB}{0, 123, 167}

\usepackage{soul}
\begin{document}


\begin{abstract} 
Coarse-grained models have proven helpful for simulating complex systems over long timescales to provide molecular insights into various processes. Methodologies for systematic parameterization of the underlying energy function, or force field that describes the interactions among different components of the system are of great interest for ensuring simulation accuracy. We present a new method, potential contrasting, to enable efficient learning of force fields that can accurately reproduce the conformational distribution produced with all-atom simulations. Potential contrasting generalizes the noise contrastive estimation method with umbrella sampling to better learn the complex energy landscape of molecular systems. When applied to the Trp-cage protein, we found that the technique produces force fields that thoroughly capture the thermodynamics of the folding process despite the use of only $\alpha$-Carbons in the coarse-grained model. We further showed that potential contrasting could be applied over large datasets that combine the conformational ensembles of many proteins to ensure the transferability of coarse-grained force fields. We anticipate potential contrasting to be a powerful tool for building general-purpose coarse-grained force fields.
\end{abstract}

\clearpage
\newpage
\section{INTRODUCTION}
Coarse-grained (CG) molecular dynamics simulations are computationally efficient and can simulate long time scale processes that are not accessible to all-atom simulations \cite{Saunders2013Coarse-GrainingBiology,Kmiecik2016Coarse-GrainedApplications,Nielsen2004CoarseMaterials,dePablo2011Coarse-GrainedNanocomposites}. They are widely used for understanding dynamical processes in physics, chemistry and biology\cite{Souza2021MartiniDynamics,Dignon2018SequenceModel,BEST2017,Joseph2021Physics-drivenAccuracy,Hyeon2011CapturingModels,Denesyuk2013Coarse-GrainedThermodynamics,Farr2021NucleosomeInteractions,Grime2016Coarse-grainedSelf-assemblyb,Mansbach2017Coarse-GrainedPeptides,Shmilovich2020DiscoverySimulation,Latham2021,Latham2019ImprovingData,Giulio2021}. The accuracy of these simulations depends on how well the force fields can describe the interactions among various components of the system under investigation. Therefore, algorithms and methodologies that can produce high-quality coarse-grained force fields (CGFF), or CG potential energy, are of key interest.

Numerous approaches have been introduced for systematically parameterizing CGFFs.\cite{Noid2013Perspective:Systems,Gkeka2020MachineSystems}.
Top-down approaches often rely on a set of experimental structural or thermodynamic properties to fine-tune CGFFs and ensure the physical relevance of CG simulations
\cite{Noid2013Perspective:Systems,Latham2022UnifyingProteins,Souza2021MartiniDynamics,Nielsen2004CoarseMaterials,Shelley2001ASimulations,Davtyan2012AWSEM-MD:Biasing,Wu2018AWSEM-IDP:Proteins,Darre2015SIRAH:Electrostatics}. On the other hand, bottom-up approaches learn CGFFs from an ensemble of atomistic configurations collected using simulations performed at finer resolution, typically with all-atom force fields \cite{Ercolessi1994InteratomicMethod,Izvekov2004EffectiveForce-matching,Zhang2018DeePCG:Networks,Zhang2018DeepMechanics}. 
From the configurational ensemble, various physical quantities and correlation functions can be computed to serve as targets for recreation with CGFF. \cite{Tschop1998SimulationPolycarbonates,Louis2000CanColloids,Davtyan2012AWSEM-MD:Biasing,Wu2018AWSEM-IDP:Proteins,Savelyev2009MolecularDNA,Akkermans2000AMelts,Ruhle2009VersatileApplications,Noid2013SystematicProtocols,Toth2007EffectivePotentials,Li2010Characterizing-Hairpin,Hansen_2005,Liwo2001Cumulant-basedField,Clark2012ThermodynamicLiquids,Mirzoev2013MagiC:Modeling}
In addition, CGFFs can also be optimized to enforce the statistical consistency between their corresponding Boltzmann distributions and the reference configurational distribution with variational methods.\cite{Izvekov2005ASystems,Shell2008TheProblems,Noid2008TheModels,Noid2008TheModelsb}. The consistency is achieved when the CG potential energy matches the potential of mean force dictated by the all-atom force field and the mapping that connects atomistic and CG configurations.

Existing variational methods optimize force field parameters by formulating and solving regression problems or maximizing the likelihood of observing the reference configurations. The force matching method\cite{Izvekov2005ASystems,Ercolessi1994InteratomicMethod} and its generalization \cite{Mullinax2009GeneralizedSystems,Kohler2022Force-matchingForces,Mullinax2009GeneralizedSystems} belong to the former category and aim to minimize the difference between forces for CG coordinates calculated from the CGFF and target values estimated from all-atom simulations. A perfect match in forces ensures that the CG energy function reproduces the potential of mean force. On the other hand, the relative entropy method\cite{Shell2008TheProblems}, or equivalently maximum likelihood \cite{Noid2013SystematicProtocols}, directly optimizes the CG energy function by minimizing the relative entropy and maximizing the overlap between the CG Boltzmann distribution and the configurational distribution from all-atom simulations. 
The relative entropy is minimized when the CG energy function reproduces the potential of mean force, and the CG Boltzmann distribution assigns high probabilities to configurations from all-atom simulations.

While existing force field parameterization methods have found great success in many applications, they are not without limitations. For example, the relative entropy method needs to run simulations to sample from trial CG potentials in every optimization step and can be computationally expensive. While the force matching method can learn the CG potential directly without iterative sampling, it often requires extra atomic force information, and the quality of the resulting potential can be sensitive to the accumulation of errors through the integration of the estimated force.

Here we developed a new variational method called potential contrasting for learning CGFFs, and applied it for multi-scale coarse-graining of protein folding. Potential contrasting generalizes the noise contrastive estimation method \cite{Gutmann2010Noise-contrastiveModels} to formulate force field parameterization into a classification problem. Input for the method is a target ensemble of protein conformations from all-atom simulations, and no atomic force information is required. When applied to the peptide Trp-cage, we found that potential contrasting can produce force fields that accurately reproduce the all-atom conformational ensemble and capture the complex folding landscape. The method also revealed the importance of including many-body potentials in CG models to describe protein biophysics with a reduced degree of freedom and implicit solvation.  In addition, we showed that potential contrasting is computationally efficient and trivially parallelizable, enabling the parameterization of transferable force fields using large datasets collected from multiple proteins.

\section{METHODS}

Potential constrasting combines a machine learning method called noise contrastive estimation\cite{Gutmann2010Noise-contrastiveModels} (NCE) with molecular simulation techniques. In this section, we first introduce NCE using the M\"uller potential\cite{Muller1979LocationProcedure} as an example. Then we present how the NCE method is generalized and used in potential constrasting to learn CGFFs for protein folding.

\subsection{Noise contrastive estimation}

\begin{figure}[t]
  \includegraphics[width=\textwidth]{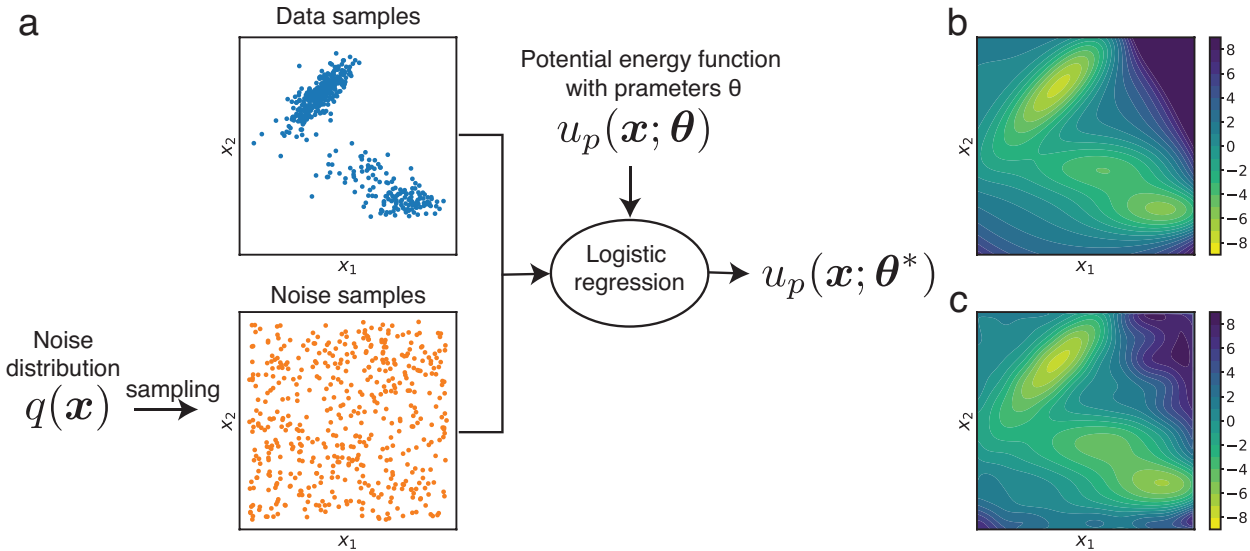}
  \caption{ 
  Noise contrastive learning accurately reproduces the M\"uller potential from sampled data only.
  (a) Illustration of the noise contrastive estimation method. Representative data samples from Monte Carlo sampling of the M\"uller potential and noise samples from a noise distribution $q(\bm{x})$ are shown in blue and orange, respectively. The target data distribution is parameterized using a potential energy function $u_p(\bm{x};\bm{\theta})$, i.e., $p(\bm{x}; \bm{\theta}) \propto \exp(-\beta u_p(\bm{x}; \bm{\theta}))$, where $\bm{\theta}$ is a set of parameters. $\bm{\theta}$ is optimized in a logistic regression to classify the data and noise samples. 
  (b) Contour plot of the M\"uller potential. Energy is shown in the units of $k_BT$.
  (c) Contour plot of the potential energy function $u_p(\bm{x}; \bm{\theta}^*)$ learned using noise contrastive estimation. Energy is shown in the units of $k_BT$.
  }
  \label{fig:mp}
\end{figure}

The NCE method\cite{Gutmann2010Noise-contrastiveModels} learns a probabilistic model on observed data. It is especially useful for learning unnormalized statistical models where the probability density function is only specified up to a normalization constant. It is evident that NCE is connected to bottom-up force field optimization, which aims to parameterize an energy function or an unnormalized Boltzmann distribution from data produced by all-atom simulations.

Here we use the M\"uller potential as an example to show how NCE helps to learn energy functions. Given a set of data (Figure \ref{fig:mp}a) drawn from the M\"uller potential with Markov chain Monte Carlo sampling, NCE aims to approximate their probability distribution with $p(\bm{x};\bm{\theta})$ defined as $\log p(\bm{x}; \bm{\theta}) = - \beta [u_p(\bm{x}; \bm{\theta}) - F_p]$, where $u_p(\bm{x}; \bm{\theta})$ is the potential energy parameterized with $\bm{\theta}$ and $F_p$ is the free energy. To optimize the parameters $\bm{\theta}$, NCE performs a logistic regression to discriminate the $N_p$ data samples $\{\bm{x}^i_p\}_{i=1}^{N_p}$ from $N_q$ noise samples $\{\bm{x}^i_q\}_{i=1}^{N_q}$  (Figure \ref{fig:mp}a) that are drawn from a noise distribution $q(\bm{x})$. Specifically, we assign binary labels of $y = 1$ and $y = 0$ to data and noise samples, respectively. NCE parameterizes the energy function by maximizing the following averaged log-likelihood of labels:
\begin{equation}  
  \label{eqn:log-likelihood}
  \ell(\bm{\theta}, F_p) = \frac{1}{N_p} \Big[ \sum_{i=1}^{N_p}\log P(y=1|\bm{x}_p^i) +  \sum_{i=1}^{N_q}\log P(y=0|\bm{x}_q^i) \Big],
\end{equation}
with 
\begin{equation}  
  \label{eqn:posterior}
  P(y = 1 | \bm{x}) = \frac{p(\bm{x}; \bm{\theta})}{p(\bm{x};\bm{\theta}) + \nu q(\bm{x})}\ \mathrm{and} \ P(y = 0 | \bm{x}) = \frac{\nu q(\bm{x})}{p(\bm{x};\bm{\theta}) + \nu q(\bm{x})},
\end{equation}
where $\nu = P(y=0)/P(y=1) = N_q/N_p$.

By definition, maximizing the above objective function forces the probability function $p(\bm{x};\bm{\theta})$ to assign high values to data samples (the first term) and low values to noise samples (the second term). In that regard, NCE is similar to the standard maximum likelihood estimation \cite{Myung2003TutorialEstimation.}, which assigns high probability on training data. Previous works\cite{Gutmann2010Noise-contrastiveModels} have proven that the solution $\bm{\theta}^*$ for optimizing $\ell(\bm{\theta}, F_p)$ behaves like the maximum likelihood estimator for large noise sample sizes and $p(\bm{x};\bm{\theta}^*)$ converges to the true data distribution. The advantage of NCE over maximum likelihood estimation is that the free energy $F_p$ is treated as a free parameter, and the optimization avoids the computationally expensive procedure for evaluating $F_p$ rigorously. In addition, a nice property of $\ell(\bm{\theta}, F_p)$ is that it is a concave function and has a unique maximum point if the potential energy function $u_p(\bm{x};\bm{\theta})$ is linear to $\bm{\theta}$.

Treating $F_p$ as an independent variable, while being advantageous, also introduces a dependence of NCE's performance on the noise distribution because the noise sample size is always limited in practice. If $p(\bm{x}; \bm{\theta})$ is a normalized density with conserved probability mass, as in the maximum likelihood optimization, increasing its value on data samples would implicitly decrease its value on regions outside the data. Such a balance of probability density is not guaranteed in NCE since $p(\bm{x}; \bm{\theta})$ is not strictly normalized due to the approximate treatment of $F_p$. The use of a noise distribution remedies this issue by allowing an explicit probability minimization for the region covered by noise samples. While a comprehensive theory is still missing on designing optimal noise distributions\cite{Chehab2022TheThink}, we find that a useful guiding principle is to design the noise distribution such that it covers the phase space occupied by and surrounding the data samples. Without significant overlap between data and noise samples, the objective function, $\ell(\bm{\theta}, F_p)$, can be trivially optimized by assigning high probability on data samples and low probability on noise samples without forcing $p(\bm{x}; \bm{\theta})$ to capture the distributional structure within the data samples. In such cases, both terms in the objective function approach the constant zero, and the gradient on $\bm{\theta}$ vanishes, hindering the optimization.

We parameterized the potential energy function $u_p(\bm{x}; \bm{\theta})$ using a two dimensional cubic spline\cite{Hastie2009ThePrediction} with 169 spline coefficients. The noise distribution $q(\bm{x})$ was chosen as the uniform distribution. $500,000$ samples were generated for both data and noise. We learned the parameters $\bm{\theta}$ by maximizing the NCE objective function $\ell(\bm{\theta}, F)$ (Eq. \ref{eqn:log-likelihood}) using the L-BFGS algorithm\cite{Zhu1997AlgorithmOptimization} (In practice, we minimize the negative of the NCE objective function). As shown in Figure \ref{fig:mp}c,  $u_p(\bm{x}; \bm{\theta}^*)$  closely matches the underlying M\"uller potential (Figure \ref{fig:mp}b), supporting the effectiveness of NCE for learning potential energy functions.

\subsection{Potential contrasting for learning force fields}

\begin{figure}[t]
  \includegraphics[width=1.0\textwidth]{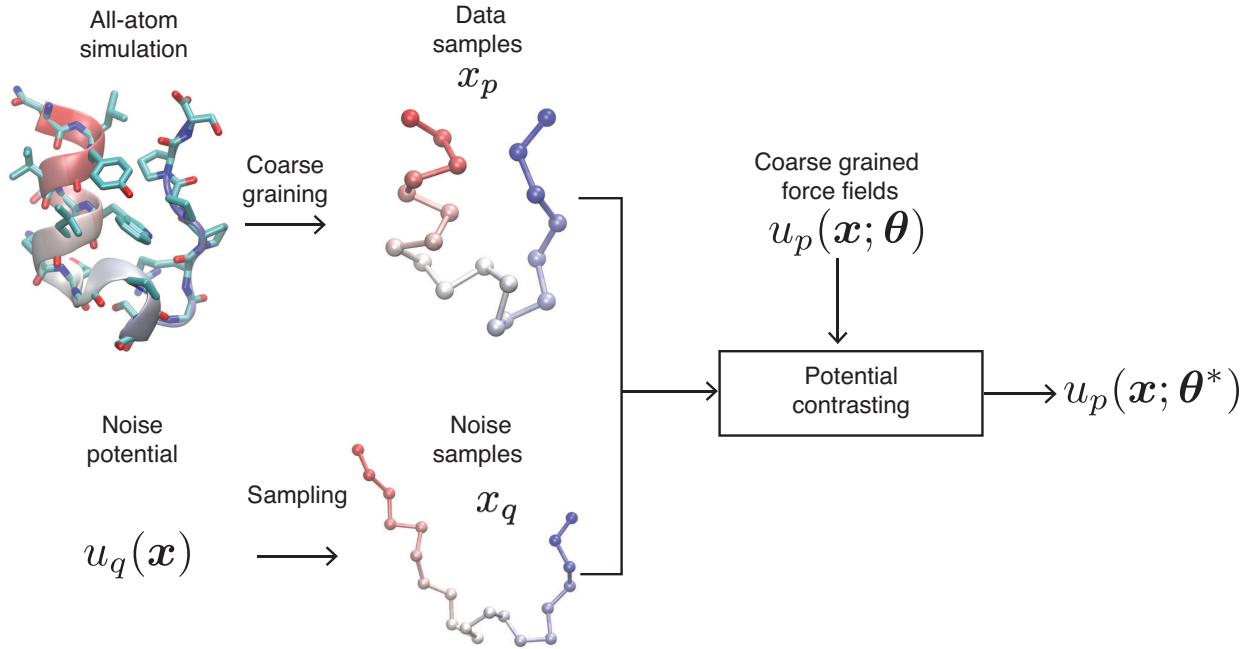}
  \caption{ 
  Workflow of the potential contrasting method for learning coarse-grained force fields for the Trp-cage protein.
  The functional form of the potential energy function is  chosen as $u_p(\bm{x}; \bm{\theta})$, where $\bm{\theta}$ is the set of parameters that need to be learned.
  The ensemble of conformations from all-atom simulations are converted into a coarse-grained ensemble using a predefined CG mapping as data samples.
  Here we map each amino acid into one coarse-grained particle at the $\mathrm{C}_\alpha$ position.
  Based on the data samples, a noise potential $u_q(\bm{x})$ is designed and used to generate an ensemble of noise conformations and optimize the parameters $\bm{\theta}$ with potential contrasting.   }
  \label{fig:pc}
\end{figure}

 We find that the current formulation of NCE, although theoretically sound, is not effective for learning molecular force fields in practice. Therefore, we developed a new method named potential contrasting by generalizing NCE and introducing a customized way of defining the noise distribution. We present details of the method with applications to protein molecules in mind, for which the development of CGFFs is of great significance but has been challenging \cite{Kmiecik2016Coarse-GrainedApplications, Latham2022UnifyingProteins}. However, potential contrasting is general and can be applied to other types of molecules.

\textbf{Generalizing NCE to unnormalized noise distributions.}
Current formulation of NCE\cite{Gutmann2010Noise-contrastiveModels} requires specifying noise distributions for which the normalized probability density can be determined at ease. This requirement restricts the choice of noise distributions because the normalization constant is difficult to compute for many probabilistic functions, including Boltzmann distributions defined by complex potentials. Here we propose that this requirement is not necessary, and generalize NCE to use noise distributions specified with a potential energy function $u_q(\bm{x})$. Specifically, we set $q(\bm{x}) = e^{-\beta[u_q(\bm{x}) - F_q]}$, where $F_q$ is the free energy.  Similarly to $F_p$, we treat $F_q$ as an extra parameter in the optimization instead of computing its value explicitly. As a result, the logistic regression objective function in Eq.~\ref{eqn:log-likelihood} becomes 
\begin{align}
  \label{eqn:glog-likelihood_0}
  \ell(\bm{\theta}, F_p, F_q) &= \frac{1}{N_p} \Big[ \sum_{i=1}^{N_p}\log \frac{1}{1 + \nu e^{-\beta[u_q(\bm{x}^i_p) - u_p(\bm{x}^i_p; \bm{\theta}) + F_p - F_q]}} +  \sum_{i=1}^{N_q}\log \frac{1}{1 + \nu^{-1} e^{-\beta[u_p(\bm{x}^i_q; \bm{\theta}) - u_q(\bm{x}^i_q) + F_q - F_p]}} \Big].
\end{align}
Because the value of $\ell(\bm{\theta}, F_p, F_q)$ in Eq. \ref{eqn:glog-likelihood_0} only depends on $\bm{\theta}$ and the difference between $F_p$ and $F_q$, we merge the two free energy into one free parameter $\Delta F = F_p - F_q$, i.e., 
\begin{align}  
  \label{eqn:glog-likelihood}
  \ell(\bm{\theta}, \Delta F) = \frac{1}{N_p} \Big[ \sum_{i=1}^{N_p}\log \frac{1}{1 + \nu e^{-\beta[u_q(\bm{x}^i_p) - u_p(\bm{x}^i_p; \bm{\theta}) + \Delta F]}} +  \sum_{i=1}^{N_q}\log \frac{1}{1 + \nu^{-1} e^{-\beta[u_p(\bm{x}^i_q; \bm{\theta}) - u_q(\bm{x}^i_q) - \Delta F]}} \Big].
\end{align}
Potential contrasting uses $\ell(\bm{\theta}, \Delta F)$ as the objective function and optimizes the parameters $\bm{\theta}$ by maximizing $\ell(\bm{\theta}, \Delta F)$. $\bm{\theta}^*$ is used to represent optimized parameters.

\textbf{Defining the noise distribution for learning CGFFs of protein folding.} 
As mentioned before, the performance of NCE depends critically on the noise distribution, which should produce samples with sufficient overlap with the training data. For low dimensional systems, a feasible choice for the noise is the uniform distribution used in the M\"uller potential example. For complex systems such as protein molecules, uniform distributions suffer the dimensionality curse to cover the relevant phase space. Our generalization to unnormalized Boltzmann distributions significantly broadens the choices of noise distributions to facilitate producing complex molecular structures that resemble data samples.  We further propose an umbrella sampling procedure to design noise potential energy functions and enhancer overlap between noise and data samples.

We design the noise potential energy function such that the noise samples contain both folded and unfolded structures to match the configurational ensemble from all-atom simulations. For a given protein, we start with an energy function that includes terms for bonds, angles and dihedral angles defined as
\begin{equation} 
  \label{eqn:bonded}
  u_\mathrm{bonded}(\bm{x}) = \sum_{i=1}^{L-1} \frac{1}{2}k_i (b_i - b_i^\circ)^2 + \sum_{i=1}^{L-2} S_\mathrm{angle}(a_i; \bm{c}^a_i) + \sum_{i=1}^{L-3} S_\mathrm{dihedral}(d_i; \bm{c}^d_i),
\end{equation}
where $L$ is the number of residues in the protein, and $b_i, a_i$ and $d_i$ represent the $i$th bond, angle and dihedral angle. A quadratic function is used for energies on bonds. $k_i$ and $b_i^\circ$ are the force constant and the equilibrium value for the $i$th bond. Cubic spline functions, $S_\mathrm{angle}$ and $S_\mathrm{dihedral}$, are used for energies on angles and dihedral angles. $\bm{c}^a_i$ and $\bm{c}^d_i$ are spline coefficients for the $i$th angle and dihedral angle, respectively. Using data samples, we fit each bonded energy term in $u_{\mathrm{bonded}}(\bm{x})$ independently such that it will reproduce the marginal distribution of the corresponding degree of freedom from the data samples. To generate both folded and unfolded structures for noise samples, we further carried out umbrella sampling simulations \cite{Torrie1977NonphysicalSampling,Tiwary2016} with  $u_\mathrm{bonded}(\bm{x})$ by biasing the root-mean-squared-deviation (RMSD) from the folded structure towards different values.

We combined configurations sampled from all $M$ umbrella simulations together to construct the noise ensemble. The probability distribution of the generalized ensemble can be described as  $p_\mathrm{gm}(\bm{x}) \propto \sum_{i=1}^M \exp(- \beta [u_i(\bm{x}) + v_i])$ \cite{PhysRevLett.63.1195,Kumar1992THEMethodb,Shirts2008StatisticallyStates,Bennett1976EfficientDatab,Ding2019FastEquations}, where $u_i(\bm{x})$ is the energy function used in the $i$th umbrella simulation that includes both $u_\mathrm{bonded}(\bm{x})$ and the bias function on the RMSD.  $v_i$ are adjustable energies that need to be fitted and added to the potential energy $u_i(\bm{x})$ so that the relative free energies of the $M$ states match the relative populations of structures sampled from these states. Correspondingly, the noise potential function can be computed as $u_q(\bm{x}) = - \beta^{-1} \log \sum_{i=1}^M \exp(- \beta [u_i(\bm{x}) + v_i])$. More details on the procedure are included in the Supporting Information.

\textbf{Extending potential contrasting to multiple proteins.}
With the developments outlined above, potential contrasting can be used to parameterize CG energy functions for a specific protein by optimizing $\ell(\bm{\theta}, \Delta F)$ defined in Eq.~\ref{eqn:glog-likelihood}.   It can be further generalized to learn CG potential functions with transferable parameters. Suppose that we can produce data and noise samples for a collection of proteins, the objective function to ensure that the CGFF reproduces the target configurational distribution for each protein can be defined as
\begin{align}  
  \label{eqn:glog-likelihood-tot}
  \ell_\mathrm{tot}(\bm{\theta}, \{\Delta F_k\}_{k=1}^K) = \sum_{k=1}^K \frac{1}{N^k_p} \Big[ &\sum_{i=1}^{N^k_p}\log \frac{1}{1 + \nu_k e^{-\beta[u^k_q(\bm{x}^{ki}_p) - u_p(\bm{x}^{ki}_p; \bm{\theta}) + \Delta F_k]}} \nonumber \\
  +  &\sum_{i=1}^{N^k_q}\log \frac{1}{1 + \nu_k^{-1} e^{-\beta[u_p(\bm{x}^{ki}_q; \bm{\theta}) - u^k_q(\bm{x}^{ki}_q) - \Delta F_k]}} \Big].
\end{align}
The above expression is a sum of potential contrasting objective functions (Eq. \ref{eqn:glog-likelihood})  introduced for each individual protein. $\{\bm{x}_p^{ki}: i = 1, \cdot\cdot\cdot N_p^k\}$ and $\{\bm{x}_q^{ki}: i = 1, \cdot\cdot\cdot, N_q^k\}$ represent the data and noise samples for the $k$th protein, with $N_p^k$ and $N_q^k$ corresponding to the respective sample sizes, and $\nu_k = N_q^k/N_p^k$. While the same energy function $u_p(\bm{x}^{ki}_p; \bm{\theta})$ with transferable parameters $\bm{\theta}$ is used, different noise potential energy functions, $ u^k_q(\bm{x}^{ki}_q)$, can be introduced for individual proteins. The aggregated objective function maintains the property of being concave if the CG energy function is linear to $\bm{\theta}$. We note that the objective function can be generalized straightforwardly if the CGFF introduces non-transferable parameters across proteins, as detailed in the Supporting Information.

\section{RESULTS}

    Potential contrasting is a general-purpose method for force field parameterization. We focus on its application to protein folding and show that it can be used to optimize CGFFs for a specific protein and a collection of proteins. Given a sufficiently flexible functional form, the force field produced by potential contrasting can accurately reproduce the configurational distribution of all-atom simulations. We further demonstrate its efficiency by simultaneously optimizing over 12 proteins to derive CG potential functions with transferable parameters.

\subsection{Coarse-grained force field for the Trp-cage protein}

We applied potential contrasting to learn CGFFs for a 20 amino acids long peptide, Trp-cage. As detailed in the \emph{Methods Section}, potential contrasting parameterizes the force field by maximizing its effectiveness in differentiating data samples from noise samples. We use as data samples a total of $N_p = 1,044,000$ conformations from a 208-$\mu$s long molecular dynamics simulation with explicit solvents performed in Ref.~\citenum{Kresten2011HowFold}. This fully atomistic simulation captures multiple folding and unfolding events for the peptide. We generated $N_q = 1,044,000$ noise samples (Figure S3) that include both folded and disordered configurations and computed the noise potential $u_q(\bm{x})$ using the umbrella sampling procedure described in the \emph{Methods Section}. In the following, we use potential contrasting to learn three CGFFs with different flexibility and complexity. For simplicity, we only use $\mathrm{C}_\alpha$ atoms to represent protein conformations and define energies, but potential contrasting can be easily generalized to more refined structural models.

\begin{figure}[t!]
  \includegraphics[width=0.95\textwidth]{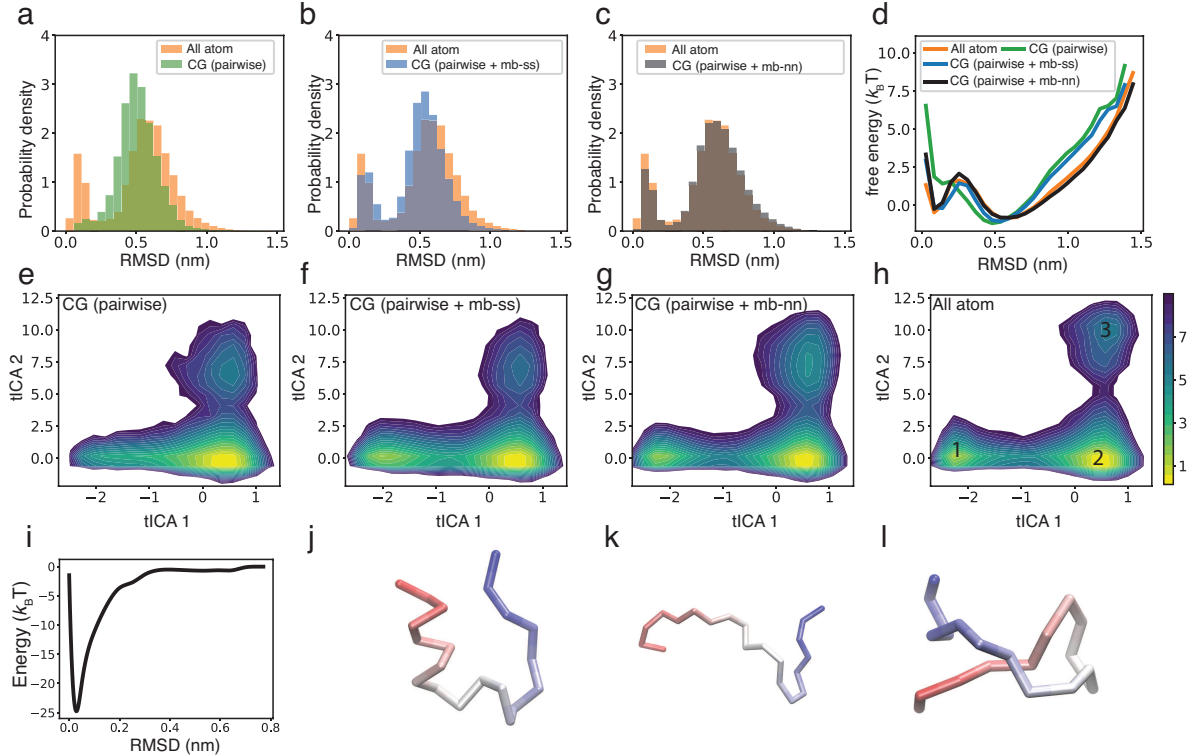}
  \caption{ Parameterizing CGFFs for the Trp-cage protein using potential contrasting and all-atom simulations.
  (a-c) Distributions of RMSD with respect to the folded structure for conformations sampled from the all-atom simulation (orange) and CG simulations with learned CG potentials that differ in the representation of the non-bonded interactions (Eq.~7-9).
  (d) Free energy profiles along the RMSD with respect to the folded structure for conformations sampled from the all-atom simulation and CG simulations with the three different learned potentials.
  (e-h) Free energy surfaces over the first two tICA coordinates for the all-atom simulation (h) and CG simulations with the three different learned potentials.
  The three meta-stable states in h are labels as 1, 2, and 3, with the corresponding representative structures shown in part j, k, and l.
  (i) The many-body potential $u_\mathrm{ss}^\mathrm{mb}(\bm{x}; \bm{\phi}^*)$ as a function of the RMSD with respect to the folded $\alpha$-helix structure.
  }
  \label{fig:trp-cage}
\end{figure}

\textbf{CGFF with bonded terms and pairwise non-bonded interactions.}
We first learned a CGFF, $u_p^\mathrm{pair}(\bm{x};\bm{\theta})$, that includes bonded terms and pairwise non-bonded terms defined as
\begin{align} 
  \label{eqn:pair}
  u_p^\mathrm{pair}(\bm{x}; \bm{\theta}) = & u_\mathrm{bond}(\bm{x}) + u_\mathrm{angle}(\bm{x}) + u_\mathrm{dihedral}(\bm{x}) + u_\mathrm{elec}(\bm{x}) + u_\mathrm{contact}(\bm{x}) \nonumber \\ 
  = &\sum_{i=1}^{L-1} \frac{1}{2} k_i (b_i - b_i^\circ)^2 + \sum_{i=1}^{L-2} S_\mathrm{angle}(a_i; \bm{c}^a_i) + \sum_{i=1}^{L-3} S_\mathrm{dihedral}(d_i; \bm{c}^d_i) + \nonumber \\
  & \sum_{i=1}^{L-4} \sum_{j=i+4}^L  \frac{q_i q_j}{4 \pi \epsilon r_{ij}} \exp(-r_{ij}/\lambda_D) + \sum_{i=1}^{L-4} \sum_{j=i+4}^L S_\mathrm{contact}(r_{ij}; \bm{c}_{ij}).
\end{align}
The bond, angle, and dihedral terms are similarly defined as in Eq.~\ref{eqn:bonded}. Non-bonded terms include electrostatics $u_\mathrm{elec}(\bm{x})$ and a contact energy term $u_\mathrm{contact}(\bm{x})$, both of which act between pairs of CG particles that are separated by four or more bonds. The electrostatic interaction is modeled using the Debye-H\"uckel theory, where $q_i$ is the net charge of the $i$th residue, $\lambda_D$ is the Debye screening length, and $r_{ij}$ is the distance between residues $i$ and $j$. The non-bonded contact energy is defined with cubic spline functions $S_\mathrm{contact}(r_{ij}; \bm{c}_{ij})$ and $\bm{c}_{ij}$ are spline basis coefficients (Figure S1). Because bond energies are much stronger than others, the parameters $b_i^\circ$ and $k_i$ were directly fitted based on the mean and the variance of the $i$th bond's distribution in the data samples. Therefore, the parameter $\bm{\theta}$ only includes spline basis coefficients, i.e., $\bm{\theta}=\{\bm{c}_i^a, \bm{c}_i^d, \bm{c}_{ij}\}$. To prevent overfitting, regularization terms on the potential energy $u_p(x; \bm{\theta})$ are added in the optimization to control their smoothness. Details on regularization terms are included in the Supporting Information. Since the energy function depends on the parameters $\bm{\theta}$ linearly, potential contrasting is guaranteed to produce a unique solution $\bm{\theta}^*$.

We carried out molecular dynamics simulations (see the Supporting Information for details) with the learned CGFF $u_p^\mathrm{pair}(\bm{x}; \bm{\theta}^*)$  to evaluate the resulting structural ensemble. Similar to that from the all-atom simulation, the distribution of RMSD with respect to the folded structure for CGFF is bimodal (Figure \ref{fig:trp-cage}a). Therefore, the learned CG potential function $u_p^\mathrm{pair}(\bm{x}; \bm{\theta}^*)$ captures both folded and unfolded structures. However, a significant discrepancy exists between the two distributions. The CG simulation produced fewer folded structures, and the two maximums of the corresponding RMSD distribution do not exactly match that of the all-atom result. The discrepancy is more clear if we convert the RMSD distribution histogram into free energy surfaces (Figure \ref{fig:trp-cage}d). Deviations can also be seen when comparing the free energy surface over the first two components of the time-independent component analysis\cite{Molgedey1994SeparationCorrelations,Naritomi2011SlowMotions,Scherer2015PyEMMAModelsb} (tICA), which describe the slowest processes observed in the simulation. The all-atom surface has three meta-stable states: one folded state and two different unfolded states that cannot be differentiated using RMSD alone (Figure \ref{fig:trp-cage}h). Although the CG simulation samples all three meta-stable states (Figure \ref{fig:trp-cage}e), it produces a smaller population of the folded state and does not capture the cooperative transitions between folded and unfolded structures (Figure S4).

\textbf{Adding many-body interactions parameterized using neural networks.}
The discrepancy between the CG and the all-atom simulations could be caused by the pair-wise potential being too restrictive and cannot capture many-body interactions that might arise due to coarse-graining. Next, we learned a more flexible energy function that includes an extra term parameterized using a feed-forward neural network with parameters $\bm{\phi}$,
i.e.,  
\begin{equation}
\label{eqn:mb_nn}
    u_p^\mathrm{nn}(\bm{x}; \bm{\theta}, \bm{\phi}) = u_p^\mathrm{pair}(\bm{x}; \bm{\theta}) + u^\mathrm{nn}_\mathrm{mb}(\bm{x}; \bm{\phi}).
\end{equation}  
The additional energy term,  $u^\mathrm{nn}_\mathrm{mb}(\bm{x}; \bm{\phi})$,  is invariant to translations and rotations and takes angles, dihedral angles, and pairwise distances as inputs (Figure S2). It can represent complex interactions involving multiple residues because the neural network is fully connected to couple different degrees of freedom\cite{Wang2020}.

A CG simulation performed with the learned potential function $u_p^\mathrm{nn}(\bm{x}; \bm{\theta}^*, \bm{\phi}^*)$ now indeed matches the all-atom results well. The maximums of the RMSD distribution are much better placed (Figure \ref{fig:trp-cage}c), suggesting that the CG simulation accurately predicts the folded structure. Importantly, the CG simulation reproduces the relative population of the folded structure and the unfolded ensemble and the free energy barrier between them (Figure \ref{fig:trp-cage}d and S4). Similarly, the free energy surface of the first two tICA coordinates (Figure \ref{fig:trp-cage}g and \ref{fig:trp-cage}h) agrees well with the all-atom one. Therefore, despite only using only $\alpha$-carbons, the CGFF captures the complex folding landscape of the peptide determined from atomistic explicit solvent simulations.

\textbf{Adding secondary structure inspired many-body potentials.}
Although parameterizing the many-body energy term using a neural network improves the accuracy of the resulting force field, it has a few disadvantages. For instance, the potential function $u_p^\mathrm{nn}(\bm{x}; \bm{\theta}, \bm{\phi})$ is not linear to $\bm{\phi}$, and the optimized parameters depends on initial conditions. Moreover, it is difficult to interpret the many-body energy in simple physical terms. To avoid these issues, we learned a CG potential function with a secondary structure based many-body energy term. Secondary structure biases are frequently incorporated into coarse-grained models as fragment memories for improved quality of structural predictions\cite{Davtyan2012AWSEM-MD:Biasing, Latham2021,Rohl2004ProteinRosetta}. They help account for cooperative effects arising from water molecules involving many residues that are challenging to describe with pair-wise potentials.
Specifically, the secondary structure based many-body energy term is defined as 
\begin{equation}
\label{eqn:mb_ss}
u_p^\mathrm{ss}(\bm{x}; \bm{\theta}, \bm{\phi}) = u_p^\mathrm{pair}(\bm{x}; \bm{\theta}) + u^\mathrm{ss}_\mathrm{mb}(\bm{x}; \bm{\phi}).
\end{equation}
It is parameterized using cubic spline functions as
$u^\mathrm{ss}_\mathrm{mb}(\bm{x}; \bm{\phi}) = S_\mathrm{ss}(\mathrm{rmsd\_ss}(\bm{x}, \bm{x}_\circ); \bm{c}^\mathrm{ss})$. Here  $\mathrm{rmsd\_ss}(\bm{x}, \bm{x}_\circ)$ is the RMSD calculated on the $\alpha$-helix (residue 3 to residue 15) between a given structure $\bm{x}$ and the folded structure $\bm{x}_\circ$. The parameter $\bm{\phi}$ includes all the spline basis coefficients $\bm{c}^\mathrm{ss}$.
This design of the energy function in Eq.~\ref{eqn:mb_ss} further ensures linear dependence on parameters and a unique solution for force field optimization.

The CG simulation results using the learned potential $u_p^\mathrm{ss}(\bm{x}; \bm{\theta}^*, \bm{\phi}^*)$ are shown in Figure \ref{fig:trp-cage}b, \ref{fig:trp-cage}d , \ref{fig:trp-cage}f, and S4. Although the many-body energy term is restricted within the $\alpha$-helix, the CG simulation correctly reproduces the relative populations of folded and unfolded states and the free energy barrier. Its performance is almost as good as the potential with a neural network based many-body term defined over the whole protein. The learned many-body potential function $u^\mathrm{ss}_\mathrm{mb}(\bm{x}; \bm{\phi}^*)$ along the $\alpha$-helix RMSD is shown in Figure \ref{fig:trp-cage}i. It has a deep well near 0 nm and quickly approaches zero when the RMSD is larger than 0.3 nm.  Therefore, the potential only plays a significant role in stabilizing the folded structure when the $\alpha-$helix is already close to the native state. Its impact is minimal when the $\alpha$-helix adopts unfolded configurations.

\subsection{Efficient optimization of transferable force fields with data from multiple proteins}

\begin{figure}[ht!]
  \includegraphics[width=0.8\textwidth]{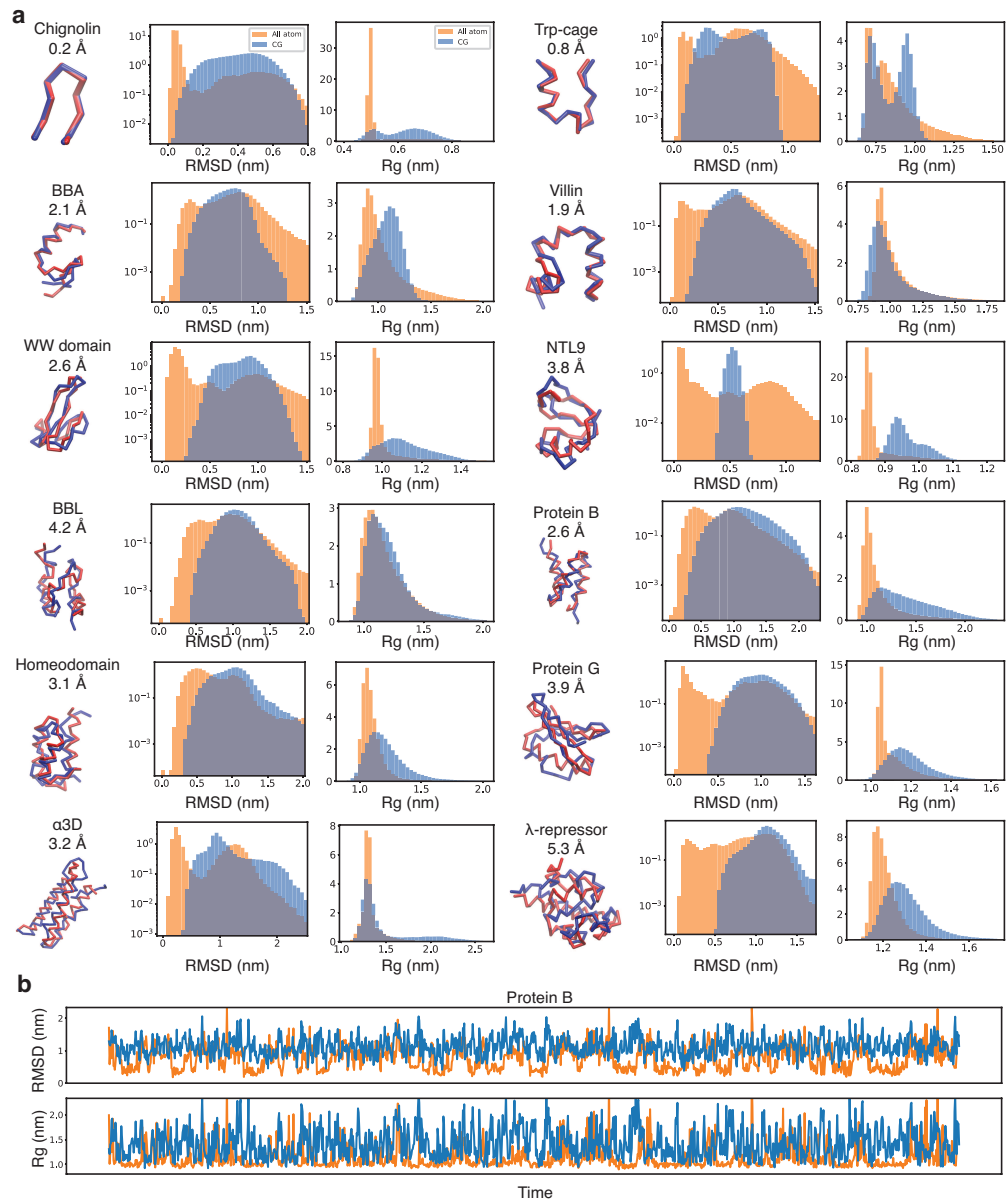}
  \caption{ 
  Comparison between all-atom simulations and CG simulations performed with the learned transferable force field.
  (a) For each of the 12 proteins, we show the folded structure (red) from the all-atom simulation, the structure (blue) from the CG simulation that has the lowest RMSD with respect to the folded structure, and the C$_\alpha$-RMSD (over all residues) between the two structures.
  The two plots on the right of structures are distributions of RMSD to the folded structure and distributions of Rg (radius of gyration) for conformations sampled from all-atom simulations (orange) and CG simulations (blue).
  (b) Trajectories of Rg and RMSD with respect to the folded structure for the all-atom simulation (orange) and the CG simulation (blue) of the Protein B. Although the data from all-atom simulations and CG simulations are plotted in the same figure, their time scales are different. Similar plots for other proteins are included in the Supporting Information.
  }
  \label{fig:all_hist}
\end{figure}

\begin{figure}[ht]
  \includegraphics[width=0.6\textwidth]{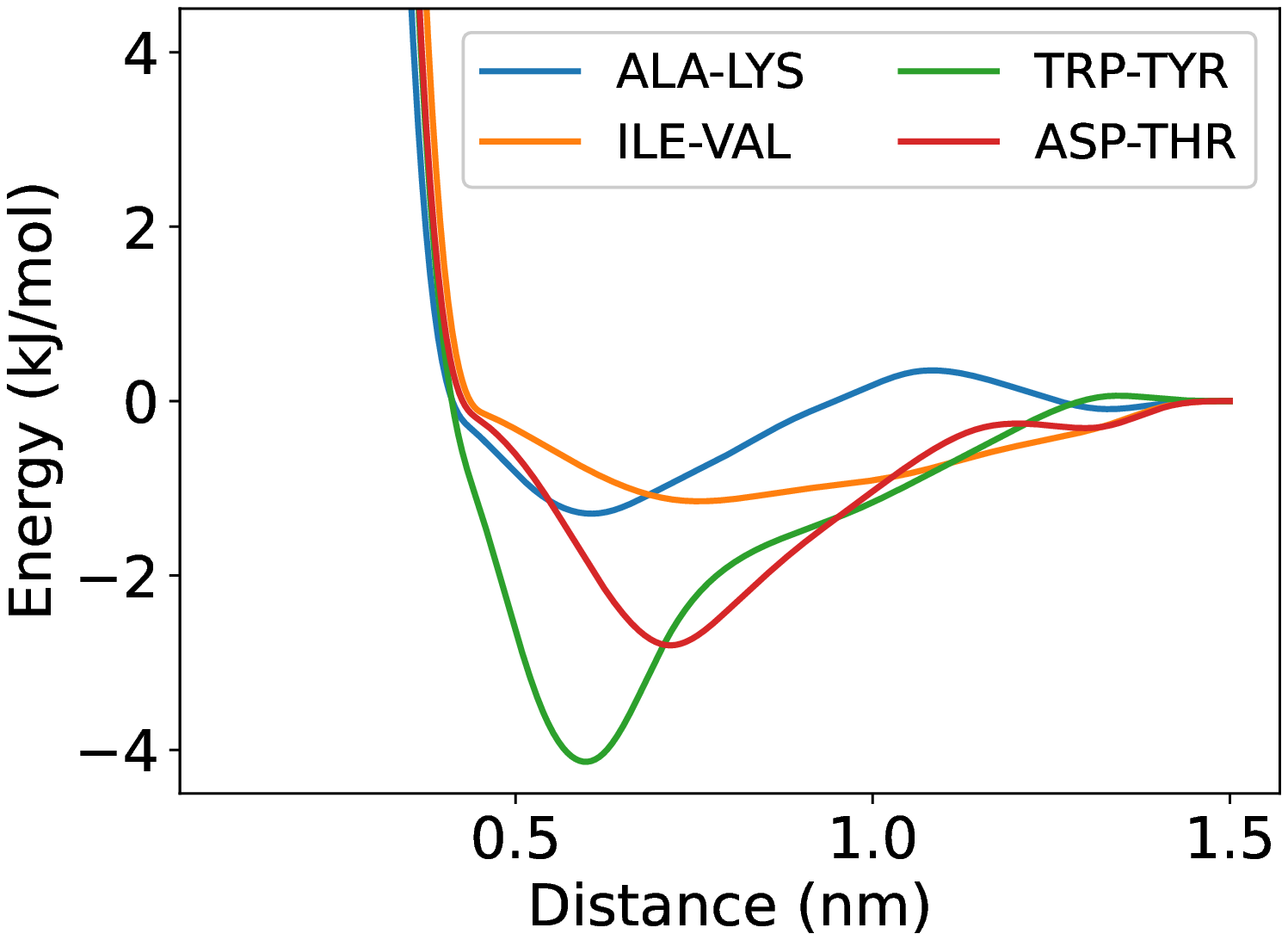}
  \caption{ 
  Learned transferable contact potential energy functions between representative pairs of amino acids.
  Similar plots for other pairs of amino acids are included in the Supplementary Information.
  }
  \label{fig:lj_contact}
\end{figure}

The above results suggest that potential contrasting is a powerful tool to parameterize flexible CGFFs for specific proteins and capture their complex folding landscapes. Next, we show that the method also allows efficient optimization of transferable force fields using all-atom simulations of 12 fast-folding proteins performed in Ref. \citenum{Kresten2011HowFold}.

The transferable force field for the $k$th protein is defined using Eq.~\ref{eqn:mb_ss} as
\begin{align} 
  \label{eqn:transferable}
  u_p^k(\bm{x}_k) = u_\mathrm{bond}(\bm{x}_k) + u_\mathrm{angle}(\bm{x}_k) + u_\mathrm{dihedral}(\bm{x}_k) + u_\mathrm{contact}(\bm{x}_k) + u_\mathrm{elec}(\bm{x}_k) + u^\mathrm{ss}_\mathrm{mb}(\bm{x}_k; \bm{\phi}_k).
\end{align}
 As a proof of principle, we only shared parameters for pair-wise non-bonded interactions and allowed protein-specific non-transferable parameters for both the bonded term and the many-body term. 
The pair-wise contact potential is now defined as
\begin{equation}
\label{eqn:contact_transferable}
u_\mathrm{contact}(\bm{x}) = \sum_{i=1}^{L-4} \sum_{j=i+4}^L S_\mathrm{contact}(r_{ij}; \bm{c}^\mathrm{contact}_{IJ}).
\end{equation}
While $S_\mathrm{contact}(r; \bm{c}^\mathrm{contact}_{IJ})$ shares the same functional form as that in Eq.~\ref{eqn:pair}, its parameters now only depend on residue types $I$ and $J$. Because $\bm{c}^\mathrm{contact}_{IJ}$ are made to depend on residue types alone, they are transferable among proteins. Our choice of limiting the force field's transferability is due to the well-known challenges of predicting secondary structures in CG models \cite{Kmiecik2016Coarse-GrainedApplications}. While potential contrasting allows efficient optimization of all parameters across proteins, the accuracy of the resulting CGFF may be poor. Allowing protein-specific potentials alleviates the challenges in describing secondary structures using CG models with only one particle per residue.

Both transferable parameters and non-transferable parameters were learned by optimizing the aggregated objective function defined in Eq.~\ref{eqn:glog-likelihood-tot}. For each of the 12 proteins, we used evenly spaced 250,000 conformations from the corresponding all-atom simulation as data samples. Using the umbrella sampling procedure described in the \emph{Methods Section}, we generated the same number of noise samples and computed the noise potentials $u_q^k(\bm{x})$. Because the energy function $u_p^k(\bm{x})$ is linear to all parameters, optimizing the aggregated objective function (Eq.~\ref{eqn:glog-likelihood-tot}) converges to a unique solution. In addition, because the aggregated objective function is a weighted sum of objective functions for individual proteins, its computing and optimization can be easily parallelized among proteins. Using 12 Nvidia Volta V100 GPUs, each assigned to calculate the potential contrasting objective function of one protein, we can optimize the aggregated objective function (Figure S5) and learn all parameters in 30 minutes.

CG simulations using the learned potential functions are compared to all-atom simulations (Figure \ref{fig:all_hist} and S6) in terms of the radius of gyration (Rg) and the RMSD from the folded structures. Structures close to the native state are sampled in the CG simulations for all proteins (Figure \ref{fig:all_hist}). The lowest RMSD for configurations sampled in CG simulations range from 0.2 \AA\ to 5.3 \AA\ and are less than 4 \AA\ for 10 out of 12 proteins. Because the CG potential functions (Eq.~\ref{eqn:transferable}) are restricted to share transferable non-bonded interactions, their performances at reproducing all-atom simulations are compromised compared to the potential function $u_p^\mathrm{ss}$ that is specific to the Trp-cage protein and has no transferable parameters. Nonetheless, the CG simulations capture folding and unfolding transitions for all but the NTL9 proteins (Figure \ref{fig:all_hist}).  The learned transferable contact potential energy functions between pairs of amino acids are shown in Figure \ref{fig:lj_contact} and S7. Although we parameterize these non-bonded contact potentials using cubic splines and do not restrict them to specific mathematical expressions, they all converge to functions that resemble the Lennard-Jones potential widely used in all-atom and CG force fields.

\section{CONCLUSION and DISCUSSION}
By generalizing noise contrastive estimation with unnormalized noise distributions, we developed a new method, potential contrasting, for learning force fields from reference molecular configurations. Potential contrasting combines the advantages of existing variational methods such as force matching and relative entropy minimization. As with the force matching method, it is computationally efficient and does not need sampling during force field optimization. Like the relative entropy method, potential contrasting does not require force information. We showed that the method is effective and succeeds in producing CG energy functions that accurately reproduce configurational distributions obtained from all-atom simulations. In addition, potential contrasting can be trivially parallelized for efficient learning of transferable CGFFs using simulation data of multiple systems. With its efficacy and efficiency, potential contrasting is well-positioned to systematically learn transferable CGFFs based on all-atom force fields, addressing one of the significant challenges in coarse-grained modeling.

Although we focused in this study on using potential contrasting to learn CGFFs, the method is general. It can be applied to learning various types of force fields. For instance, potential contrasting can be readily applied to parameterize implicit solvent models using all-atom simulations with explicit water molecules. With further development, it could also be used to improve existing all-atom force fields by incorporating information from quantum mechanical calculations or experimental data. Such applications and development will be investigated in future studies.

Our use of unnormalized noise distributions produced with umbrella sampling is essential for parameterizing accurate CGFFs. Unnormalized noise distributions defined with molecular energy functions allow the generation of noise samples that resemble the configurations produced from all-atom simulations. Therefore, significant overlap in the phase space between noise and data samples can be achieved. Such overlap can be difficult to ensure with arbitrary noise distributions since all-atom simulations only sample limited regions of phase space with low energy. We note that,  upon training molecular simulation data,  probabilistic models parameterized with normalizing flows\cite{Rezende2015VariationalFlowsc,Dinh2016DensityNVP,Papamakarios2019NormalizingInferenced} have been shown to produce realistic and stable molecular conformations \cite{Gao2020FlowModels,Noe2019BoltzmannLearningb,Ding2020ComputingModels,Ding2021DeepBAR:Computationb,Wirnsberger2020,Kohler2022Force-matchingForces}. These models have indeed been proposed to serve as noise distributions for contrastive learning to guarantee overlap with data samples \cite{Gutmann2010Noise-contrastiveModels,Gao2020FlowModels}.  However, we found that using flow-based models as noise distributions produced CGFFs with sub-par quality.  Similar findings have been reached in other recent studies as well \cite{Chehab2022TheThink}. By optimizing the overlap with data samples, flow-based models may hinder the minimization of probability for regions outside the data. Further research is needed to design optimal noise distributions in NCE.

\begin{acknowledgement}
This work was supported by the National Institutes of Health (R35GM133580).
\end{acknowledgement}

\begin{suppinfo}
Detailed procedure for generating noise samples for learning CGFF of protein folding,
learning CG potential functions with both transferable and non-transferable parameters,
cubic splines for flexible potential energy parameterization,
parameter optimization and regularization,
molecular dynamics simulations with the CGFFs,
many-body interactions parameterized using neural network,
(Figure S1) cubic spline basis,
(Figure S2) the neural network used for parameterizing the many-body energy term,
(Figure S3) distributions of RMSD with respect to the folded structure of the Trp-cage protein for the noise samples generated using umbrella sampling,
(Figure S4) trajectories of RMSD with respect to the folded structure of the Trp-cage protein for conformations from the all-atom simulation and the CG simulations,
(Figure S5) convergence of the aggregated loss function with weight decay during the optimization
of the CG potential functions that have both transferable and non-transferable parameters,
(Figure S6) trajectories of raidus of gyration and RMSD with respect to the folded structure from the
all-atom simulation and the CG simulation of all 12 proteins,
(Figure S7) learned transferable contact potential energy functions between pairs of amino
acids.
(Table S1) setup used in umbrella sampling used for generating noise samples.
\end{suppinfo}

\clearpage
\newpage
\bibliography{achemso-demo}

\end{document}